\documentstyle[aps,prl,epsfig,preprint]{revtex}
\begin{document}
\newcommand{\ic}{\;\;\; ,}
\newcommand{\ip}{\;\;\; .}

\title{A Baryon Model in Covariant Constraint Dynamics}
\author{H.W.\ Fricke and C.C.\ Noack }
\address{Institut f\"ur Theoretische Physik, Universit\"at Bremen,
         D-28334 Bremen, Germany}
\author{Submitted to Physical Review Letters \\[2ex]
       }
\date{\today}
\maketitle
\tightenlines

\begin{abstract}
An important ingredient of parton or string cascade models for
ultrarelativistic heavy-ion reactions is a parton description of the
baryon. Whereas previous models needed the concept of a diquark in an
essential way, we have developed a new model using Dirac's approach of
Poincar\'{e}-covariant many-body dynamics with constraints.

In our model, the baryon is described as a dynamical set of three
valence quarks and a fourth `particle', the ``junction'', which carries
the momentum fraction of the sea quarks as well as all of the glue.

The model's parameters are the quark current masses, and one interaction
strength, determined by the proton radius. Thus the model has no
adjustable free parameters. Nevertheless, we obtain a remarkably good
fit to the valence quark structure functions of the baryon.
\end{abstract}

\paragraph*{Introduction.}
Among the methods employed in describing ultrarelativistic heavy ion
reactions theoretically, various cascade models \cite{Wer93,Gei95},
(cf.\ also \cite{Dea91}) play an important r\^{o}le. These models follow
the space-time development of the reaction dynamics in phase space; they
are therefore essentially classical models. They simulate the nuclear
reaction by a series of elementary propagations, fusions and decays of
partons or strings. When successful, cascade models not only reproduce
the experimental data as asymptotic states of the reaction, but give
some information on the dynamical, non-equilibrium development of the
intermediate states, including a phase transition to a
quark-gluon-plasma (QGP) --- if such a phase transition should indeed
occur.

The basic objects in \cite{Wer93} are strings; (confined) baryons or
mesons are described by such strings sweeping (1+1)-dimensional periodic
hypersurfaces in space-time (``yo-yo motions''). A baryon is thus
modeled as a system of a quark and a diquark. Parton distributions are
not included in this ansatz.

In this letter we will present a new Poincar\'e-covariant dynamical
model for a baryon, leading to yo-yo-like motion of its constituents.
Three valence quarks and a ``junction'' (representing sea quarks and all
gluonic degrees of freedom), interact as classical point particles via a
quasi-potential. This model, which is free of adjustable parameters,
approximately reproduces the valence quark  structure functions. The
$Q^2$ dependency of the structure functions will be discussed at the end
of the letter.

\paragraph*{Constraint Dynamics.}
In constructing a Poincar\'e-covariant system of $N$ interacting
point-particles, one has to face the consequences of the no-interaction
theorem \cite{Sud63}. This theorem asserts that the only canonical
Hamiltonian theory that is fully covariant under Poincar\'e
transformations is that of a system of non-interacting particles.
Dirac's method of introducing dynamical constraints \cite{Dir49}
circumvents this problem.

In this paper we use a set a $(2N)$ second-class constraints, written
in the scheme of \cite{Sam82} (cf.\ also \cite{Sor89}). The phase space
is expanded to $8N$ dimensions :
\[ \Gamma := \left\{ \left. q^\mu_{(i)} , p^\mu_{(i)} \,\right|\,
             i=1\dots N, \mu=0\dots 3 \right\} \ip \]
An evolution parameter $s$ -- without any direct physical meaning -- is
introduced for the parametrization of the phase space trajectories. A
set of $(2N-1)$ Poincar\'e-covariant constraints
$ \{ \Phi_r \,|\, r=1 \dots (2N-1) \} $
and one connection $\Phi_{2N}$ between the evolution parameter $s$ and
the phase space variables are introduced which reduce the degrees of
freedom of the system to a physical $6N$-dimensional hypersurface
$\Gamma^\prime$ . We use the standard notation of writing $A \approx B$
to denote that an equation holds only on $\Gamma^\prime$; we also employ
a summation convention over equal indices whereever applicable.

The equations of motion are generated by a Poincar\'{e}-invariant
`Hamiltonian'
$H := \lambda^j \Phi_j$ via the Poisson brackets,
\begin{mathletters}   \label{dp}
\begin{eqnarray}
\frac{d q_{(i)}}{ds} \approx \left\{ q_{(i)}, H \right\}
                     \approx \left\{ q_{(i)}, \Phi_j \right\} \lambda^j
                                                  \;\; ,  \\
\frac{d p_{(i)}}{ds} \approx \left\{ p_{(i)}, H \right\}
                     \approx \left\{ p_{(i)}, \Phi_j \right\} \lambda^j
                                                  \;\; .
\end{eqnarray}
\end{mathletters}
The Lagrange parameters $\lambda^j$ are determined by the requirement
that $H$ preserve the constraints:
$ \frac{d \Phi_i}{ds} \approx \frac{\partial \Phi_i}{\partial s}
      + \left\{ \Phi_i, \Phi_j \right\} \lambda^j \approx 0 $ .
Denoting the matrix of the Poisson brackets
$ \left\{ \Phi_i, \Phi_j \right\} $ by ${\cal P}_{i,j}$ , the
$\lambda^j$ are given by
$ \lambda^j \approx -\left({{\cal P}^{-1}}\right)^{j,k}\,
             \frac{\partial \Phi_{k} }{\partial s} $ ;
since of all the constraints only $\Phi_{2N}$ depends explicitly on $s$
, we obtain
\begin{eqnarray}
\lambda^j & \approx &  -\left({{\cal P}^{-1}}\right)^{j,2N}\,
             \frac{\partial \Phi_{2N} }{\partial s}  \;\; .
\label{lambda}
\end{eqnarray}
[\,It is straightforward to see that $\lambda^{2N} \approx 0$ ;
thus $H$ is independent of $s$\,] .
Using Eqs.\ (\ref{dp}) and (\ref{lambda}), one obtains the equations of
motion for physically interpretable variables on the hypersurface
$\Gamma^\prime$.

With constraint dynamics it is possible to construct a system consisting
of two massive particles $m_{(1,2)}$ , with equations of motion
comparable to those of a string model \cite{Beh94}. It is important that
the period of this yo-yo motion is of the same order as that of the
string model:
($T_{\text{yoyo}} := 2M/\kappa_{\text{Lund}}$), because the period fixes
the maximum time for one fragmentation \cite{Wer93}. In order to achieve
this, we use a quadratic quasi-potential (rather than the linear
potential of the Lund model), which in turn increases the effective
masses of the two particles:
\begin{eqnarray*}
\left( {m_{(1,2)}}_{\text{eff}} \right)^2
      & := & \left( p_{(1,2)} \right)^2
         =   \left( m_{(1,2)} \right)^2 - V        \\
      &  = & \left( m_{(1,2)} \right)^2 - \kappa^2
             \left( q_{(1)} - q_{(2)} \right)^2  \ip
\end{eqnarray*}
The dynamical system is defined by the constraints
\begin{mathletters}
\label{IPC_2Teilchen}
\begin{eqnarray}
\Phi_{1,2} & := & \left( p_{(1,2)} \right)^2
                   - \left( m_{(1,2)} \right)^2
                   + \kappa^2 \left( q_{(1)} - q_{(2)}
                  \right)^2                                     \\
\Phi_{3~~} & := & (q_{(1)} - q_{(2)}) \cdot
                  \left( p_{(1)} + p_{(2)}\right)               \\
\Phi_{4~~} & := & \frac{p_{(1)} + p_{(2)}}{\left| p_{(1)} + p_{(2)}
                  \right|} \cdot q_{(1)} - s     \ip
\end{eqnarray}
\end{mathletters}

The physical meaning of these constraints is the following: $\Phi_{1,2}$
guarantee that the particles are on mass shell wherever the
quasi-potential vanishes. Their effective masses increase with
increasing distance . $\Phi_{3}$ ensures that in the center-of-momentum
system (CMS) the particle distance is space-like , and $\Phi_{4}$
connects the evolution parameter $s$ with the particle times in the CMS .

\begin{figure}
   \centerline{\psfig{file=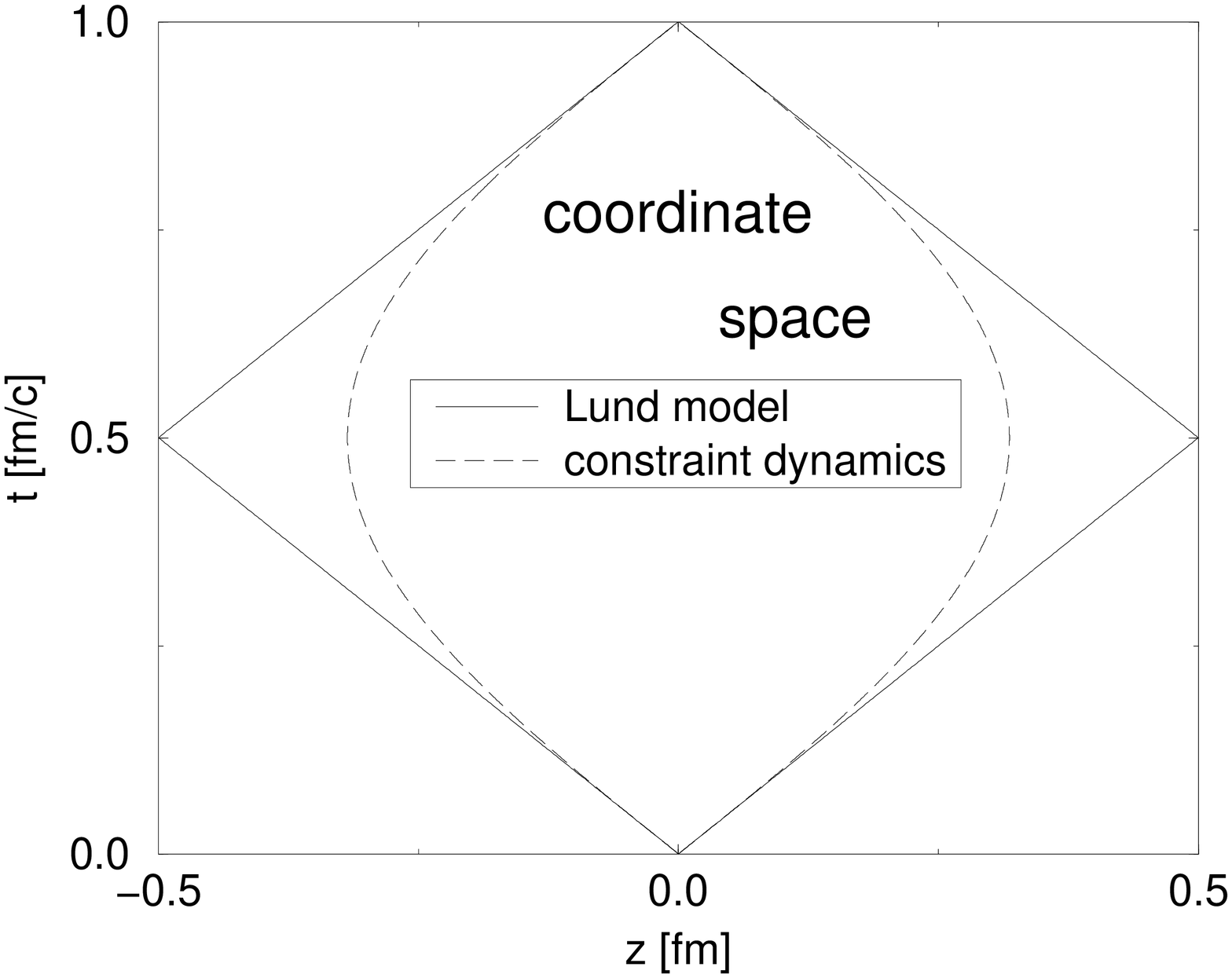,width=10cm,angle=270}}
   \centerline{\psfig{file=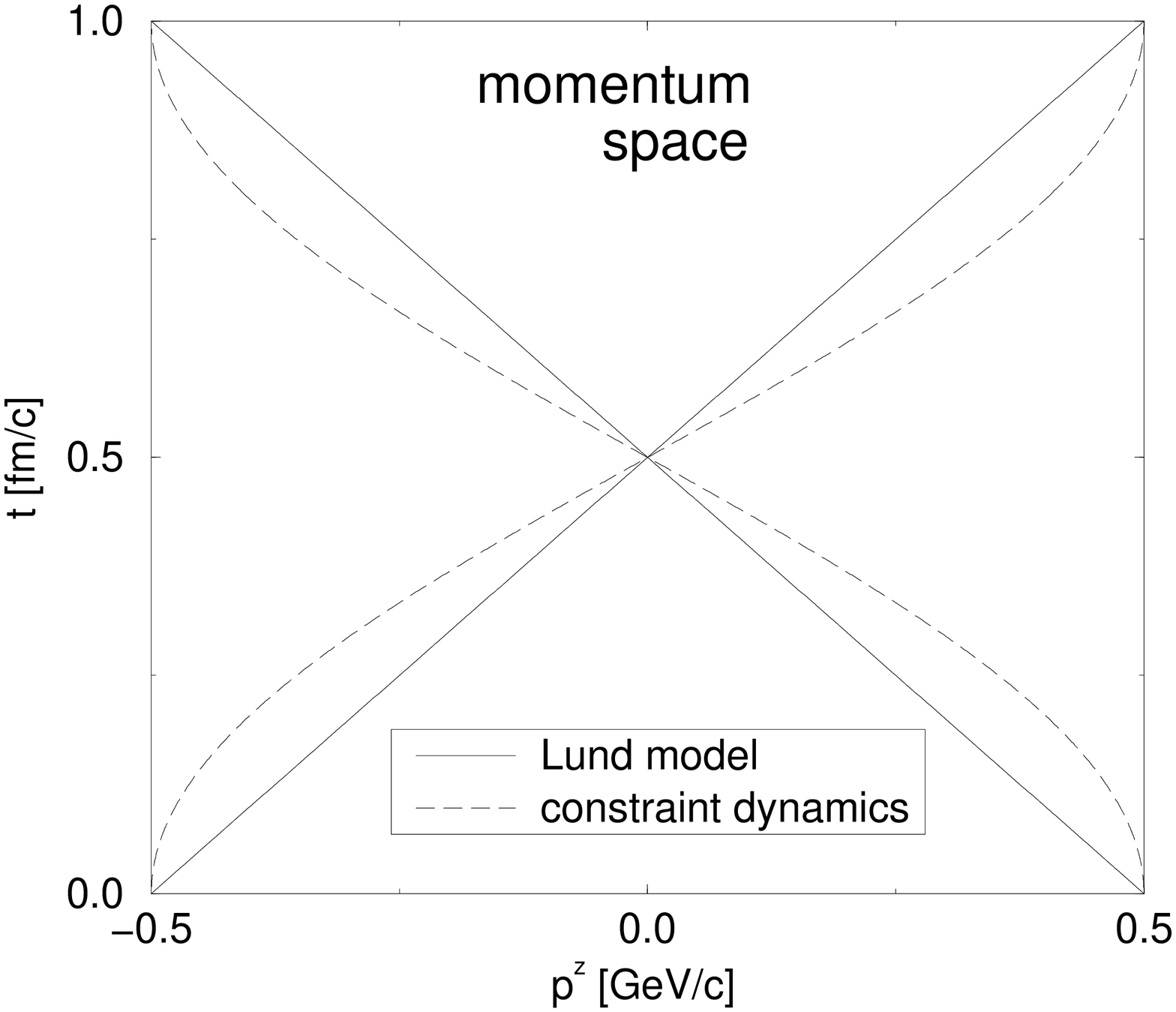,width=10cm,angle=270}}
   \caption[]{Solution of the equations of motion in the CMS of the
              constraint dynamics model
              [\,Eqs.\ (\ref{IPC_2Teilchen})\,],
              compared to the Lund solution, in coordinate and in
              momentum space. $m_{(1,2)}=0 .$
             }
   \label{Fig_1}
   \end{figure}

The solutions of the equations of motion for this system are
trigonometric functions (Fig.\ \ref{Fig_1}). In the CMS and with
vanishing particle masses, the period is $T := \pi M / (2\kappa)$ ,
where $ M^2 := \left( p_{(1)} + p_{(2)}\right)^2 $ . Setting
$\kappa = \pi/4 \kappa_{\text{Lund}}$ , one obtains the same period as
in the Lund model.

\text{$\:$} This fully Poincar\'e-covariant model can thus be thought of
as simulating a meson of total mass $M$, composed of two quarks with
current masses $m_{(1,2)}$. As in the Lund model, the quarks in the meson
are confined, and this increases their effective masses. In comparison,
the yo-yo string solution is only boost-invariant in one direction.
Current quark masses are not included.

\paragraph*{Model of the Baryon.}    \label{sec:baryon}
In the spirit of the above, a naive model for a baryon would be a system
of three valence quarks, each carrying $1/3$ of the baryon momentum.
This, however, is too gross a simplification, since from deep inelastic
scattering we know that about one half of the momentum is carried by
sea quarks and gluonic degrees of freedom. In the model to be described
now, we will introduce a fourth (classical) particle, the `junction'
\cite{Ros80,Kha96}, which binds the valence quarks, and thus
models these degrees of freedom (Cf.\ Fig.\ \ref{Fig_2}).

In keeping with the discussion in the previous section, we will again
use a quadratic quasi-potential depending on the distances between the
junction and the valence quarks [\,again, constraints will ensure that
these distances are space-like in the CMS\,] . This quasi-potential
should affect the effective mass $\left( p_{(J)} \right)^2 $ of the
junction, because pushing away a valence quark will increase the number
of sea quarks and gluons, which are modeled by the junction. For
simplicity, we will assume the valence quarks to always remain on mass
shell.

The above physical requirements are embodied in the following
constraints:

\begin{mathletters}
\label{IPC_4Teilchen}
\begin{eqnarray}
\Phi_{1,2,3} & := & \left( p_{(1,2,3)} \right)^2
                    - \left( m_{(1,2,3)} \right)^2             \\
\Phi_{J~~~~} & := & \left( p_{(J)} \right)^2  +\kappa^2 \sum_{i=1}^3
                    \left( q_{(i)} - q_{(J)} \right)^2         \\
\Phi_{5,6,7} & := & (q_{(1,2,3)} - q_{(J)}) \cdot P            \\
\Phi_{8~~~~} & := & \frac{P}{\left| P \right|} \cdot q_{(J)} - s  \ic
\end{eqnarray}
\end{mathletters}
where $P := p_{(J)} + \sum_{i=1}^3 p_{(i)}$ is the total 4-momentum of
the system.

The equations of motion for this system have no simple solution in
closed form; but they are, of course, easily solved numerically. For the
simulation of a proton, we use $M_p=\left| P \right|=0.938 \text{\,GeV}$
(proton mass), $\kappa=0.5 \text{\,GeV/fm}$ (leading to a proton radius
of $\approx 0.8 \text{\,fm}$) and estimated values for the current
masses of the quarks \cite{PDTables}: $m_{(1)}=m_{(2)}=m_{(u)}=5
\text{\,MeV}$, $m_{(3)}=m_{(d)}=10 \text{\,MeV}$.

\paragraph*{$Q^2$-independent Structure Functions.}
For fixed values of the variable $x:=Q^2/2M\nu$, the substructure of
nucleons is rather independent of the 4-momentum transfer ($Q^2$).
($M$ denotes the nucleon mass and $\nu$ the transferred energy.) In the
parton model this scaling behaviour is explained in terms of the
presence of point-like charged constituents, called `partons'. Using the
parton hypothesis, $x$ can be interpreted as the fractional momentum of a
parton in the infinite momentum frame. The distribution functions of
this variable can be obtained from the data of deep inelastic $e-p$
collision experiments. For comparison with our results, we shall use the
parametrization of the valence quark distribution functions
$u_v(x),d_v(x)$ given in \cite{EHQL} .

To obtain valence quark structure functions in our model described by
Eqs.\ (\ref{IPC_4Teilchen}), we first need the variable $x$ in the
context of our model. We define the longitudinal momentum fraction of
particle $i$ in terms of the ratio of the light-cone variables:
$ x_{(i)} :=
\frac{p^+_{(i)}}{P^+} := \frac{p^0_{(i)}+ p^z_{(i)}}{P^0 + P^z} $ .
Note that this definition is flavor-dependent because we are using
flavor-dependent quark masses (cf.\ end of the previous section).

\begin{figure}
   \centerline{\psfig{file=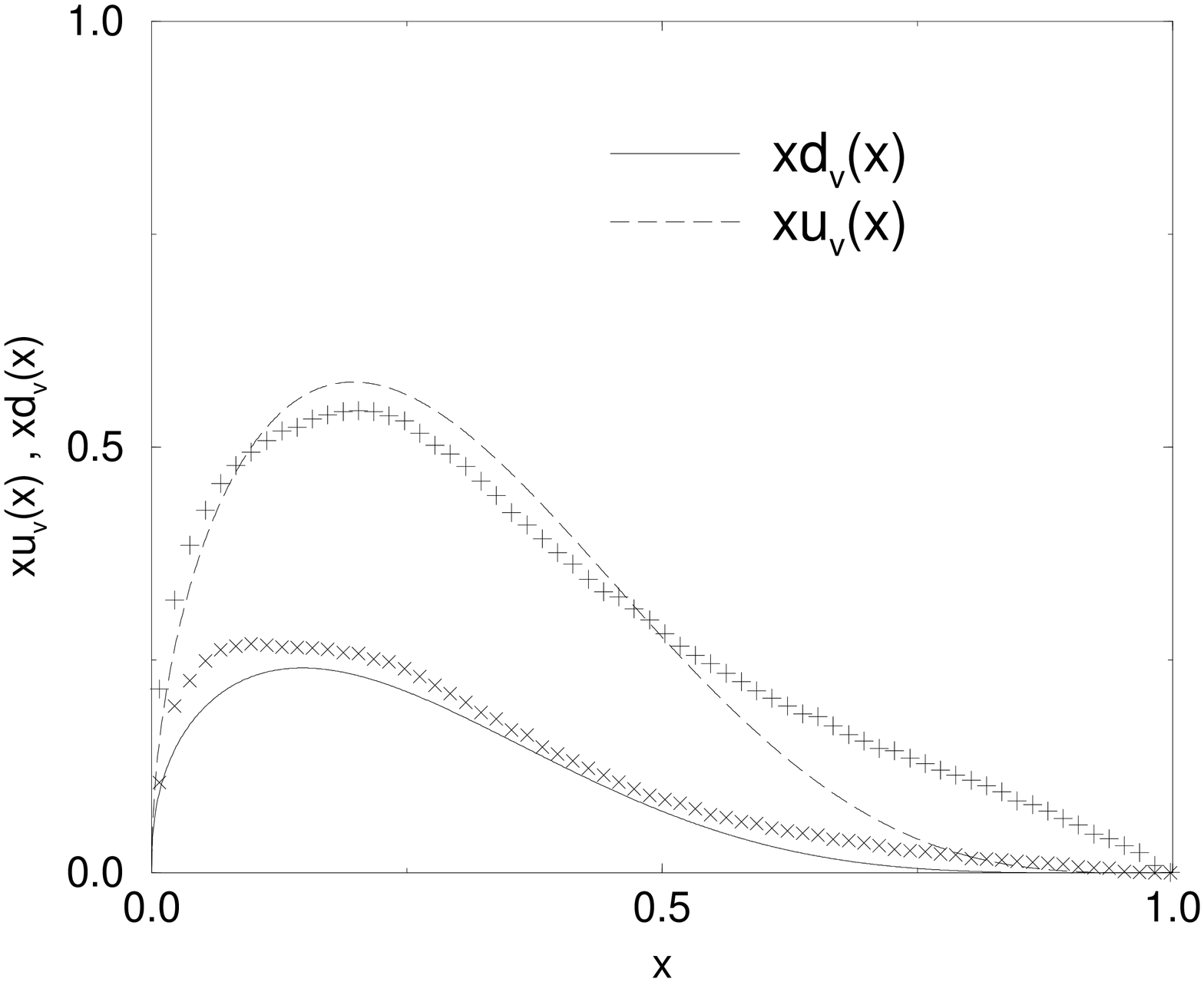,width=12cm,angle=270}}
   \caption[]{Sampled distributions of the valence quark longitudinal
              momentum fraction $x u_v(x)$ (up quark) and $x d_v(x)$
              (down quark). The solid and dashed lines are the results
              obtained with the parametrization \cite{EHQL} .
             }
   \label{Fig_3}
   \end{figure}

We have then sampled distributions of these momentum fractions by
randomly choosing different initial conditions for the dynamical
system of Eqs.\ (\ref{IPC_4Teilchen}); during the evolution of the
system with one set of initial conditions, values of $x_{(i)}$
are sampled and averaged over. The results of these calculations are
given in Fig.\ \ref{Fig_3} and compared with the parametrization of
\cite{EHQL} .

The rather good overall agreement of our result with the data seems
rather remarkable, given the fact that ours is a purely classical
particle model, and no free parameters were fitted in obtaining these
results.

The tails of these distributions are definitely overestimated by our
model. However, one should note that for $x \approx 0.8$ the number of
events in the sample is vastly less than for $x \approx 0.1$ (by about a
factor of $10^{-6}$), so any error in the tail of the distribution
hardly contributes to the momentum distribution of the proton.

A fragmentation mechanism using the baryon model described here has been
presented elsewhere \cite{DPG}.

\paragraph*{$Q^2$-dependent Structure Functions.}

Experiments at HERA with high 4-momentum transfer show that the total
momentum carried by the valence quarks decreases with increasing $Q^2$,
and that the number of gluons and sea quarks increases \cite{Glueck} .
In terms of our model, this would mean that the effective mass $\left(
p_{(J)} \right)^2 $ of the junction should increase with $Q^2$, as it is
the junction which models these degrees of freedom. However, since by
construction the momentum transfer $Q^2$ does not enter the dynamics of
our model baryon at all, the only way to include this effect is by
introducing an additional \emph{phenomenological} mass term in the
equations of motion for the junction. We thus replace the constraint
$\Phi_{J}$ in Eqs.\ (\ref{IPC_4Teilchen}) by

\begin{equation}
\Phi_{J} := \left( p_{(J)} \right)^2 - \left( m_{(J)}(Q^2) \right)^2
              + \kappa^2 \sum_{i=1}^3 \left( q_{(i)} - q_{(J)} \right)^2
                                       \label{IPC_4Teilchen_Q}  \ic
\end{equation}
with a phenomenological $Q^2$-dependent term $m_{(J)}(Q^2)$. Instead of
simply fitting $m_{(J)}(Q^2)$ to the data, we proceed as follows: The
measured parton distribution functions for valence as well as for sea
quarks and gluons, $f_k(x,Q^2)$, are parametrized in \cite{Glueck}. From
these we calculate \emph{integrated} momentum fractions
$\bar{x}_k(Q^2) := \int_0^1 x f_k(x,Q^2) dx$ . $\bar{x}_{\text{glue}}$
and $\bar{x}_{\text{sea}}$ are a measure of the amount of the
corresponding `mean field' carried by the $J$ particle in our model
and which we want to represent by $m_{(J)}$. However, since a certain
part of this `mean field' is generated dynamically by the system in the form
of the effective mass squared of the junction, $\left( p_{(J)}
\right)^2$, we have to subtract this part, to avoid double counting. We
are thus led to the ansatz
\[ m_{(J)} :=  \left[ \bar{x}_{\text{glue}} +\bar{x}_{\text{sea }}
              -\frac{\alpha}{1 -\alpha}
               \left(\bar{x}_{d_v} +\bar{x}_{u_v} \right) \right]
              \cdot M_p  \ic  \]
where the proton mass $M_p$ is needed for dimensional reasons. The
factor $\alpha$, finally, is determined by the requirement that
$m_{(J)}(Q^2)$ vanish at the point where we connect with the
$Q^2$-independent calculation: $Q^2=2\;\text{GeV}^2$.  Thus we take
$m_{(J)}(Q^2)$ to be given by

\begin{eqnarray}
m_{(J)}(Q^2) &:=& \left[ \bar{x}_{\text{glue}}(Q^2)
                         +\bar{x}_{\text{sea }}(Q^2)
                                         \right. \nonumber \\
             &  & \left. -\frac{0.55}{0.45}
                  \left(\bar{x}_{d_v}(Q^2) +\bar{x}_{u_v}(Q^2) \right)
                  \right] \cdot M_p  \ip        \label{mj}
\end{eqnarray}

\begin{figure}
   \centerline{\psfig{file=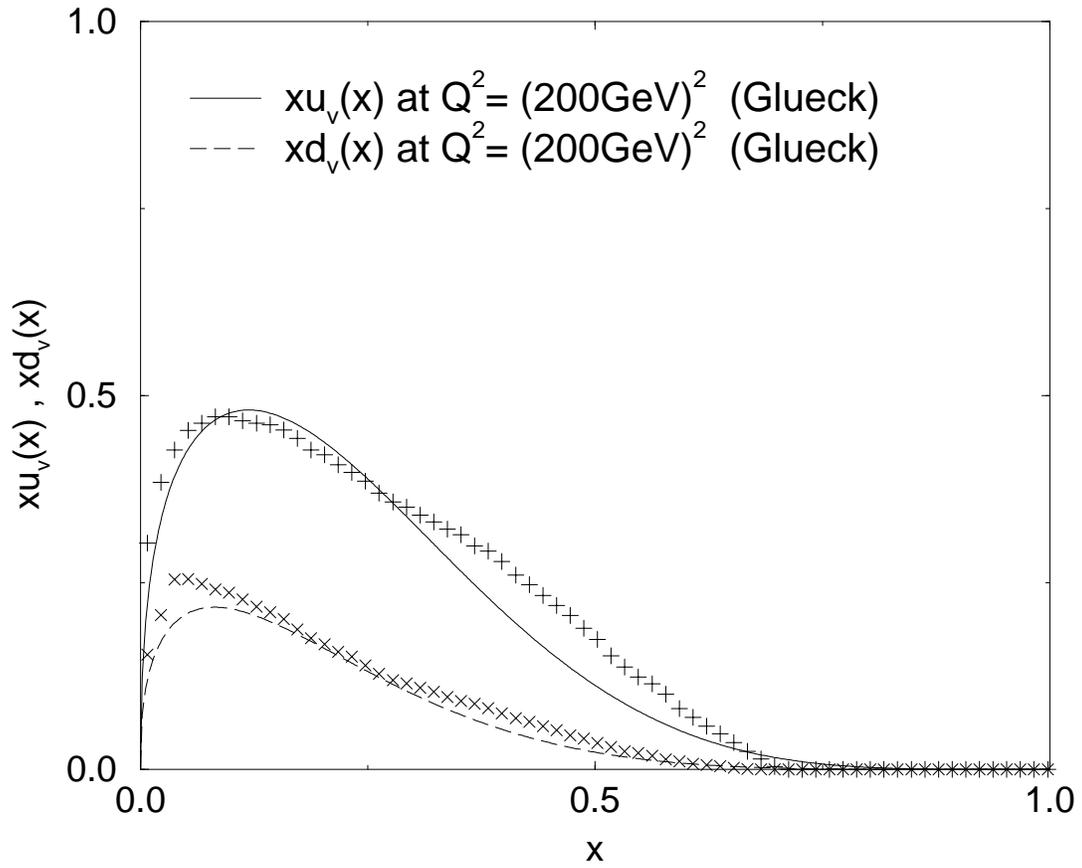,width=12cm,angle=270}}
   \caption[]{Sampled distributions of the valence quark longitudinal
              momentum fraction $x u_v(x)$ (up quark) and $x d_v(x)$
              (down quark), at $Q^2=(200\;\text{GeV})^2$. The solid and
              dashed lines are the results obtained with the
              parametrization \cite{Glueck}.
             }
   \label{Fig_4}
   \end{figure}

\begin{figure}
   \centerline{\psfig{file=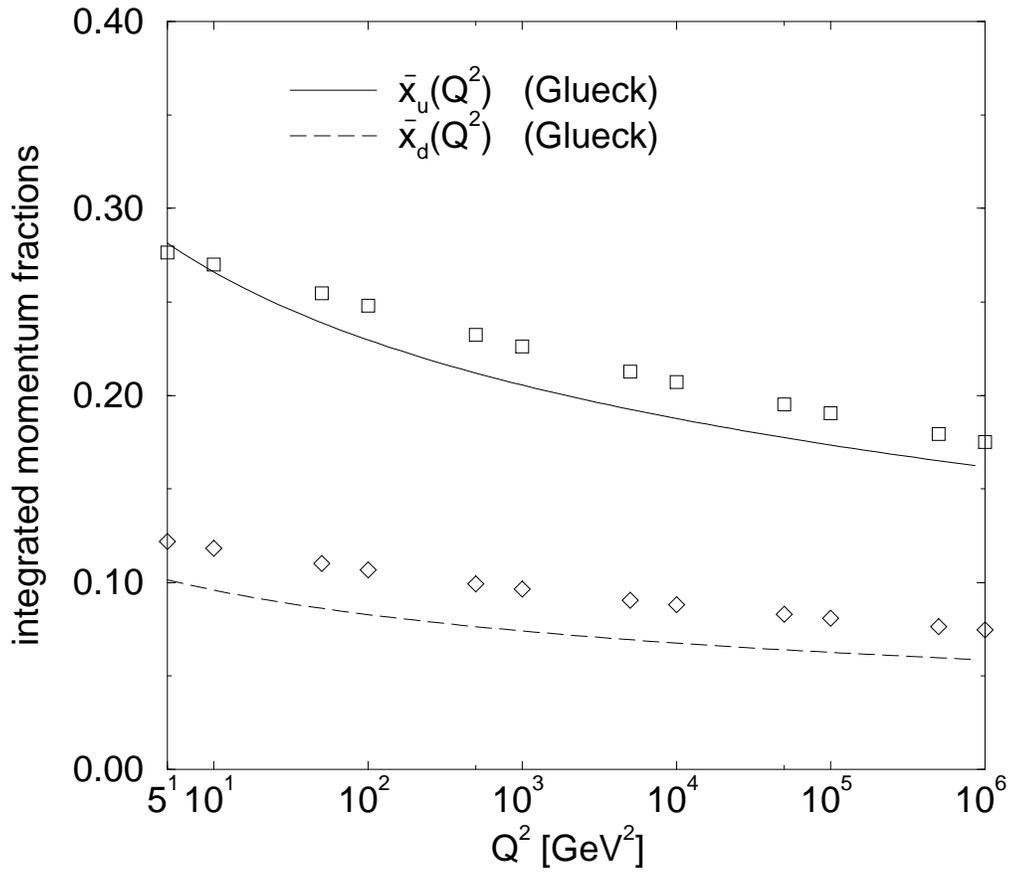,width=12cm,angle=270}}
   \caption[]{Sampled integrated momentum fractions $\bar{x}_u,
              \bar{x}_d$ of the valence up quarks ($\Box$) and down
              quarks ($\diamond$), as a function of $Q^2$. The solid and
              dashed lines are the results obtained with the
              parametrization \cite{Glueck}.
             }
   \label{Fig_5}
   \end{figure}

With this $Q^2$-dependent term added, we again solve the equations of
motion of the system and sample the distributions of the valence quark
structure functions in the manner described above. The results are
presented in Figs.\ \ref{Fig_4} and \ref{Fig_5}, in comparison with
\cite{Glueck}.

In conclusion, we have presented a truly Poincar\'{e}-covariant model
for a baryon in terms of a completely classical particle description
which, in spite of its simplicity, seems to describe some of the
features of the baryon surprisingly well. We believe this model
(possibly with further refinements) to be useful in constructing models
for ultrarelativistic heavy-ion reactions of the parton cascade type.
We are currently in the process of developing such a code.

\pagebreak

   \pagebreak
   {\begin{figure}
   \centerline{\psfig{file=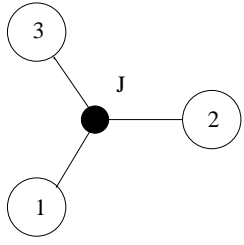,width=4cm}}
   \caption[]{Model for the baryon: a junction (J) is binding the
              valence-quarks (1,2,3) by a confinement potential. The
              junction carries the average momentum of sea quarks and
              gluonic degrees of freedom.
             }
   \label{Fig_2}
   \end{figure}}

   \pagebreak

\end{document}